\documentclass{pasa}%

\usepackage{graphicx}

\title[Non-thermal emission from the CWB Apep]{The non-thermal emission from the colliding-wind binary Apep}

\author[del Palacio et al.]{del Palacio, S,$^1$, Benaglia, P,$^1$, De Becker, M.$^2$, Bosch-Ramon, V.$^3$ and Romero, G.~E.$^{1}$ 
\affil{$^1$ Instituto Argentino de Radioastronom\'{\i}a (CONICET;CICPBA;UNLP), C.C. No 5, 1894, Villa Elisa, Argentina}
\affil{$^2$Space sciences, Technologies and Astrophysics Research (STAR) Institute, University of Li\`{e}ge, Belgium}
\affil{$^3$Departament de F\'{i}sica Qu\`antica i Astrof\'{i}sica, Institut de Ci\`encies del Cosmos (ICC), Universitat de Barcelona (IEEC-UB), Mart\'{i} i Franqu\`es 1, E08028 Barcelona, Spain}
}%

\jid{PASA}
\doi{10.1017/pas.\the\year.xxx}
\jyear{\the\year}

\usepackage{aas_macros}
\usepackage{hyperref} 
\hypersetup{colorlinks,citecolor=blue,linkcolor=blue,urlcolor=blue}

\hypersetup{draft}

\usepackage{xspace}
\newcommand{\flux}{\,erg\,s$^{-1}$\,cm$^{-2}$\xspace}	
\newcommand{\kms}{\,km\,s$^{-1}$\xspace} 
\newcommand{\Msunyr}{$\mathrm{M}_\odot$\,yr$^{-1}$\xspace} 
\newcommand{\Rsun}{$\mathrm{R}_\odot$\xspace} 

\begin{document}

\begin{frontmatter}
\maketitle

\begin{abstract}
The recently discovered massive binary system Apep is the most powerful synchrotron emitter among the known Galactic colliding-wind binaries. This makes this particular system of great interest to investigate stellar winds and the non-thermal processes associated with their shocks. This source was detected at various radio bands, and in addition the wind-collision region was resolved by means of very-long baseline interferometric observations. We use a non-thermal emission model for colliding-wind binaries to derive physical properties of this system. 
The observed morphology in the resolved maps allows us to estimate the system projection angle on the sky to be $\psi \approx 85^\circ$. The observed radio flux densities also allow us to characterise both the intrinsic synchrotron spectrum of the source and its modifications due to free--free absorption in the stellar winds at low frequencies; from this we derive mass-loss rates of the stars of $\dot{M}_\mathrm{WN} \approx 4\times10^{-5}$~\Msunyr and $\dot{M}_\mathrm{WC} \approx 2.9\times10^{-5}$~\Msunyr. 
Finally, the broadband spectral energy distribution is calculated for different combinations of the remaining free parameters, namely the intensity of the magnetic field and the injected power in non-thermal particles. We show that the degeneracy of these two parameters can be solved with observations in the high-energy domain, most likely in the hard X-rays but also possibly in $\gamma$-rays under favourable conditions.
\end{abstract}

\end{frontmatter}

\section{INTRODUCTION }
\label{sec:intro}

Massive stars launch powerful, hypersonic winds. Despite playing a key role in the evolution of stars and their feedback on the interstellar medium, the mass carried by these winds is still largely uncertain due to difficulties in the determination of their properties \citep[e.g.][]{Puls2008}. In addition, massive stars are most likely found forming binary systems, which also affects their evolution \citep{Sana2012}.

A rich phenomenology arises in massive binaries in which the stellar winds collide (dubbed \textit{colliding-wind binaries}, CWBs), generating a region of strong shocks where relativistic particles can be accelerated \citep[][]{Eichler1993}. These particles produce broadband non-thermal radiation \citep[e.g.][]{Benaglia2003, Pittard2006, Reitberger2014, delPalacio2016, Pittard2021}. So far, more than 40 CWBs have been identified as particle accelerators, mainly by means of evidence of non-thermal emission found in the radio domain \citep{DeBecker2013, DeBecker2017}. The energy budget of non-thermal particles depends on the kinetic power of the wind, which for a star with mass-loss rate $\dot{M}$ and wind velocity $v_\mathrm{w}$ is $\approx 0.5 \dot{M} \, v_\mathrm{w}^2$. In addition, particle acceleration is more efficient for fast shocks \citep[e.g.][]{Drury1983}. With high mass-loss rates and wind velocities, Wolf--Rayet (WR) stars produce some of the most powerful shocks in this type of sources, although only a couple of CWBs made up by two WR stars are known \citep{Rosslowe2015}\footnote{Updated catalogue of WR stars: \url{http://pacrowther.staff.shef.ac.uk/WRcat/}}.

Multi-frequency radio continuum observations are a unique tool to investigate the physical processes undergoing in a source. In the case of CWBs, their radio emission has two contributions: a thermal one from the individual stellar winds, and a non-thermal one from the wind-collision region (WCR). On the one hand, the free--free emission from the individual stellar winds is steady and typically presents a flux density depending on frequency ($\nu$) as $S_\nu \propto \nu^{0.6}$ \citep{Wright1975}. On the other hand, the synchrotron emission from the WCR is likely to be modulated with the orbital period of the system. In addition, its spectral index is (intrinsically) negative, with a canonical value of $-0.5$. However, this emission has to travel through the ionised stellar winds before reaching us. The free--free opacity to this radiation is frequency-dependent, being higher at lower frequencies. This can change drastically the observed spectrum, making the spectral index to be less negative or even positive \citep[e.g.][]{Dougherty2003}. The significance of the absorption depends on the square of the density of the absorbing medium (in this case, the stellar winds), and therefore this effect is more relevant in massive CWBs with dense stellar winds \citep{delPalacio2016}. In conclusion, a complete sampling at low frequencies of the spectral energy distribution (SED) of the emission is required to investigate the interplay between absorption and emission processes in the stellar winds and the WCR. 

The CWB Apep is a peculiar case of a CWB made up of two WR stars \citep{Callingham2019}. Recently, \cite{Marcote2021} were capable of resolving its WCR using very long baseline interferometric observations (VLBI). This proved unambiguously the presence of relativistic electrons accelerated at the shocks and provided constraints on the system parameters. In particular, \cite{Marcote2021} obtained the wind momentum rate ratio from the shape and position of the WCR. In this work we aim to model the SED and emission morphology of the CWB Apep in order to better constrain the mass-loss rate of the stellar winds. The precise determination of $\dot{M}$ also allows us to reduce the uncertainties about the kinetic power of the stellar winds, and therefore in the efficiency at which the wind kinetic energy is converted into non-thermal particle energy through diffusive acceleration at the shocks. We also provide predictions of the high-energy emission from this system that are important for future observational campaigns in the X-ray and $\gamma$-ray energy bands.

\section{TARGET AND OBSERVATIONS}
\label{sec:apep}

\subsection{The colliding-wind binary Apep}

This WN + WC system, located at a distance of $2.4^{+0.2}_{-0.5}$~kpc ($RA = 16$:00:50.48, $DEC=-51$:42:45.4; J2000), was identified as a peculiar case of a CWB made up of two WR stars by \cite{Callingham2019}. The system presents a spectacular dust plume that adopts a pinwheel shape on large scales (hundreds of AU), most likely due to the orbital motion of the stars. In a following study, \cite{Callingham2020} derived the stellar wind velocities. The stars are separated by a projected distance $D_\mathrm{proj} \approx 47$~mas \citep{Han2020}, which leads to a projected linear distance of 113~AU. In addition, \cite{Marcote2021} presented VLBI observations with the Australian Long Baseline Array at 2.2~GHz in which the WCR could be resolved. The shape and position of the WCR allowed them to derive a wind momentum rate ratio of $\eta = 0.44\pm0.08$, which implies a mass-loss rate ratio of $\dot{M}_\mathrm{WC}/\dot{M}_\mathrm{WN} = 0.73 \pm 0.15$. However, it is not possible to derive the individual mass-loss rates from this relation alone. \cite{Marcote2021} considered typical values of $\dot{M}$ for WN stars within $(2-10)\times10^{-5}$~M$_\odot$\,yr$^{-1}$ and adopted a reference value of $\dot{M}_\mathrm{WN} = 5\times10^{-5}$~M$_\odot$\,yr$^{-1}$, although this value has a large uncertainty. 

The known values of various parameters of the system are compiled in Table~\ref{tab:parameters}. In case that some specific value is unknown, we assume the typical value that corresponds to the given spectral type of the star.

\begin{table*}
    \caption{Parameters of the system Apep adopted in this work. Values marked with $\dagger$ were obtained in this work as described in the text.}
    \centering
    \begin{tabular}{@{}lll@{}}
    \hline\hline
    Parameter 		            &   Value			            &	Reference		\\
    \hline%
    Distance			        &	$d=2.4^{+0.2}_{-0.5}$~kpc		            &   \cite{Callingham2019} \\ 
    Projected system separation	&   $D_\mathrm{proj}=47\pm6$~mas    &   \cite{Han2020} \\
    Projection angle$\dagger$   &   $\psi=85^\circ$             &   This work (Sec.~\ref{sec:model_maps}) \\
    Wind momentum rate ratio	        &	$\eta=0.44\pm0.08$	                &   \cite{Marcote2021}	\\
    \hline
    Stellar temperature         &	$T_\mathrm{eff,WN}=65\,000$~K &  Typical \citep[e.g.][]{Crowther2007, Hamann2019}	\\
    Stellar radius  		    &	$R_\mathrm{WN} = 6$~\Rsun		&     Typical \citep[e.g.][]{Hamann2019}\\    
    Wind terminal velocity      &	$v_{\infty,\mathrm{WN}} = 3500\pm100$~\kms 	&   \cite{Callingham2020} \\ 
    Wind mass-loss rate$\dagger$&	$\dot{M_\mathrm{WN}} = 4\times10^{-5}$~\Msunyr & This work (Sec.~\ref{sec:model_radio})	\\ 
    Wind mean atomic weight     &	$\mu_\mathrm{WN} = 2.0$ 		        &	Typical \citep[e.g.][]{Leitherer1995}	\\ 
    Wind temperature            &	$T_\mathrm{w,\mathrm{WN}} = 0.3\,T_\mathrm{eff,\mathrm{WN}}$ & Typical \citep[e.g.][]{Drew1990} \\ 
    Wind filling factor         &	$f_\mathrm{WN} = 0.2$	                &   Typical \citep[e.g.][]{Runacres2002}	\\
    \hline
    Stellar temperature         &	$T_\mathrm{eff,\mathrm{WC}}=60\,000$~K & Typical \citep[e.g.][]{Crowther2007, Sander2019}	\\
    Stellar radius  		    &	$R_\mathrm{WC} = 6.3$~\Rsun		&      Typical \citep[e.g.][]{Sander2019}            \\    
    Wind terminal velocity      &	$v_{\infty,\mathrm{WC}} = 2100\pm200$~\kms 	&	\cite{Callingham2020}		\\ 
    Wind mass-loss rate$\dagger$&	$\dot{M_\mathrm{WC}} = 2.9\times10^{-5}$~\Msunyr & This work (Sec.~\ref{sec:model_radio})	\\ 
    Wind mean atomic weight     &	$\mu_\mathrm{WC} = 4.0$ 		        &	Typical \citep[e.g.][]{Cappa2004}	\\ 
    Wind temperature            &	$T_\mathrm{w,\mathrm{WC}} = 0.3\,T_\mathrm{eff,\mathrm{WC}}$ 	&	Typical \citep[e.g.][]{Drew1990} \\ 
    Wind volume filling factor         &	$f_\mathrm{WC} = 0.2$	                &   Typical \citep[e.g.][]{Runacres2002}	\\ 
    \hline\hline
    \end{tabular}
    \label{tab:parameters}
\end{table*}

\subsection{Observations in the radio band} \label{sec:radio_obs}
The system was observed with ATCA at 1.4~GHz  revealing the brightest CWB with a flux density of $S_{1.4} = 166 \pm 15$~mJy \citep{Callingham2019}. The additional flux density measurement at 19.7~GHz, $S_{19.7} = 27.9 \pm 0.9$~mJy, allowed \cite{Callingham2019} to derive a negative spectral index of $\alpha = -0.71 \pm 0.05$. Additionally, \cite{Marcote2021} presented VLBI observations at 2.2~GHz; they reported a flux density measurement of 60~mJy, though this is likely underestimated due to the VLBI observations resolving out structure of the extended WCR. In any case, these observations proved unambiguously the presence of relativistic electrons accelerated at the WCR and the synchrotron nature of the radiation they produce. 
More recently, \cite{Bloot2021} presented a more complete dataset at radiofrequencies comprising observations with the uGMRT (at 255 and 583~MHz), ASKAP (at 887.5~MHz) and ATCA (at 1--3~GHz). These observations revealed that the source is variable on timescales of several years. Nonetheless, in this work we do not aim to explore the radio variability due to the large uncertainties in the orbital parameters of Apep. For this reason, we only include in our forthcoming analysis the ATCA observations conducted in May 2017, and the uGMRT and ASKAP data obtained between 2018--2020 summarised in \cite{Bloot2021}. We also assume an additional $10\%$ systematic error in flux density values due to calibration uncertainties.

\begin{figure}[htpb]
\includegraphics[width=\linewidth]{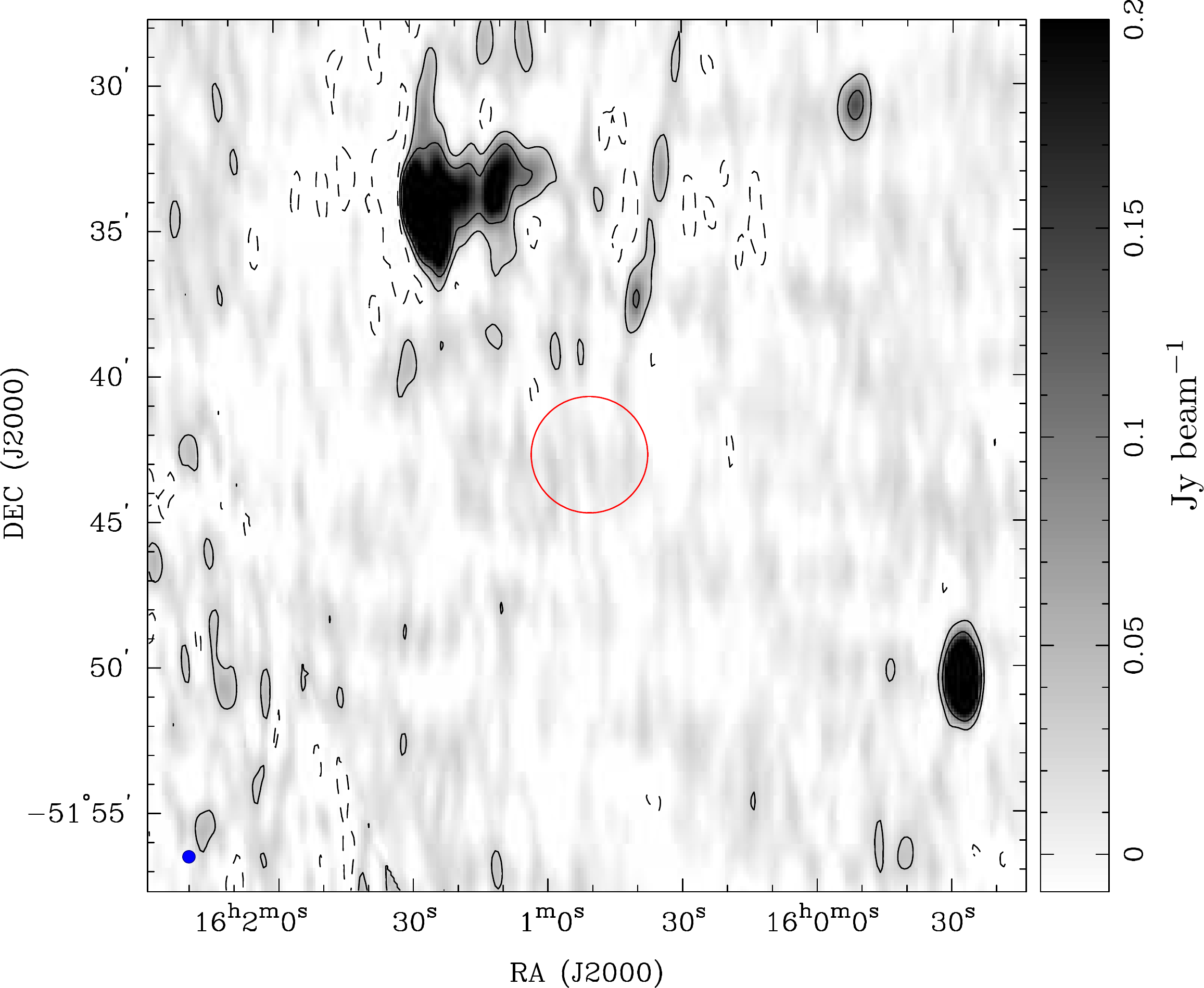}
\caption{Cutout from the GMRT 150~MHz all-sky radio survey \citep{Intema2017}. The position of Apep is highlighted with a red circle. The synthesised beam is $25''\times25''$ and is shown in the bottom left corner. Contour levels at $-30$, 30 and 100~mJy\,beam$^{-1}$ are also shown.}
\label{fig:apep_gmrt}
\end{figure}

In this work, we also report on the analysis of the radio emission at 150~MHz from the TIFR GMRT Sky Survey Alternative Data Release 1\footnote{http://tgssadr.strw.leidenuniv.nl/doku.php} \citep[TGSS ADR1;][]{Intema2017}. These data were taken between 2010--2012, so they are not simultaneous with the remaining dataset but also not too distant in time. The corresponding cutout is shown in Fig.~\ref{fig:apep_gmrt}. We analyse the 25''-synthesised-beam image with the {\sc miriad} software package \citep{Sault1995}.
The source Apep is not detected at this frequency; such non-detection yields an upper limit of $S_{\rm 0.15GHz} \leq 10$~mJy at a 1-$\sigma$ level.

\subsection{Observations in the X-ray band}\label{sec:x-ray_obs}
\cite{Callingham2019} analysed the X-ray data taken with the satellites \textit{XMM-Newton} and \textit{Chandra}; in these observations the system is completely unresolved, and therefore the observed spectrum is the sum of the individual stellar components and the WCR. They reported that the source is not variable and that the spectrum is predominantly thermal, as can be inferred by the presence of a strong Fe line. The spectrum is also significantly absorbed below 2~keV.

In this work we are interested in studying the non-thermal emission originated in the WCR. With this end, we further investigate the possibility of constraining a putative power-law component in the hard X-ray spectrum, as can be expected for a bright CWB \citep[e.g.][]{Hamaguchi2018, delPalacio2020}. Given that the source is not significantly variable, we choose to analyse the \textit{XMM-Newton} observation ObsID~0201500101, for which the source is pointed closest to on-axis and is observed with all EPIC cameras. 
We reduce the observation and extract the source spectrum using \text{SAS V19.0.0}, and we use the software \texttt{XSPEC V12.11.1} \citep{Arnaud1996} for the spectral fitting. We focus our analysis on the spectrum $>3$~keV, where 
the putative power-law component is more relevant. 

Following \cite{Callingham2019}, we fit the spectrum with an \textit{apec} model that is suitable for calculating the thermal emission of a high-temperature plasma. The absorption is modelled using a \textit{tbabs} component and setting abundances to \cite{Wilms2000} values. We obtain a total flux in the 3--10~keV band of $F_{3-10\,\mathrm{keV}} = (6.4\pm0.1)\times10^{-12}$~\flux. However, this value is only a loose upper limit of the actual non-thermal X-ray emission. Taking into account the emission from the relativistic particles in the WCR, the spectrum should be modelled as a combination of a thermal and a non-thermal (power-law) component, such as \textit{tbabs * (apec + po)} \citep[c.f.][]{delPalacio2020}. For the power-law component we fix the photon spectral index to the expected value of $\Gamma = 1 - \alpha = 1.71$ (see forthcoming Sec.~\ref{sec:model}); unfortunately, the normalisation of this component is poorly constrained by the data. 
We add a \textit{cflux} multiplicative component to the power-law component to calculate its flux, obtaining a value of $F_{3-10\,\mathrm{keV}} \approx 0.8^{+0.9}_{-0.8} \times10^{-12}$~\flux. However, the addition of the power-law component does not improve the quality of the fit and has to be taken with great caution; in fact, the lower limit of the retrieved flux is consistent with zero. Nonetheless, we can obtain a stringent 1-$\sigma$ upper limit of the flux of the power law component of $F_{3-10\,\mathrm{keV}} < 1.7 \times10^{-12}$~\flux.
%

\section{EMISSION MODEL}
\label{sec:model}

We use the multi-zone model described in \cite{delPalacio2016} to calculate the non-thermal emission from the WCR. This model is suitable for adiabatic and quasi-stationary shocks with a laminar flow, as expected for systems separated by several AU such as Apep \citep[e.g.][]{Pittard2009}. Moreover, this extended model incorporates consistently the transport of relativistic particles along the shocks and the emission they produce at each location in a 3D space. This allows us to properly correct the emission for position-dependent absorption along the line of sight. Below we present a brief summary of the model, and we refer the reader to \cite{delPalacio2016, delPalacio2020} for further details.

The stars are separated by a linear distance \mbox{$D=d\,D_\mathrm{proj}/\sin{\psi}$}, where $\psi$ is the projection angle on the sky. We model the WCR at scales of the binary system separation, where the effects of orbital motion do not affect significantly its shape. The WCR structure is then treated as an axi-symmetric surface under a thin shocked shell approximation. The thermodynamic quantities at each shocked shell (one for each stellar wind) are calculated using analytical prescriptions. In particular, the magnetic field pressure is parameterised as a fraction $\eta_B$ of the thermal pressure at each position. Relativistic particles are assumed to accelerate in the WCR region and flow together with the shocked fluid. As they stream, particles cool down due to different processes and produce broadband radiation. This intrinsic emission is then corrected for absorption in the local matter and radiation fields. In Appendix~\ref{appendix:dist} we present a slight modification on how the particle energy distribution is calculated with respect to the base model in \cite{delPalacio2020}.

The relativistic particle distribution injected at a given position in the WCR is a power law with the spectral index directly given by the radio observations through $p=-2\alpha + 1 = 2.42$. This assumption is reasonable because radio emitting-electrons do not have time to cool, which would modify the electron energy distribution, and absorption effects are not significant at frequencies above 1.4~GHz (see below). The normalisation of this distribution is such that the injected power is a fraction $f_\mathrm{NT}$ of the total power available for particle acceleration \citep[which is only a fraction of the total power of the stellar winds;][]{delPalacio2016}. This power is distributed in electrons and protons as $f_\mathrm{NT} = f_\mathrm{NT,e} + f_\mathrm{NT,p}$. We adopt a parameterisation $f_\mathrm{NT,e} = K_\mathrm{e,p} f_\mathrm{NT}$, with \mbox{$K_\mathrm{e,p} = 0.02$} \citep[e.g.][]{Merten2017}.

The more distinct signatures of the non-thermal emission can be found in the extremes of the energy spectrum: at low radio frequencies and at high-energy X-rays and $\gamma$-rays. On the one hand, the emission in the low-frequency radio band is produced by the synchrotron mechanism. This radiation can be significantly attenuated by free--free absorption (FFA) in the ionised stellar winds. We consider an increased free--free opacity due to clumping in the stellar winds by a factor $f^{-1/2}$, where $f\approx0.2$ is the volume filling factor of the wind \citep[e.g.][]{Runacres2002}. On the other hand, the non-thermal X-ray emission is produced by anisotropic inverse Compton (IC) up-scattering of stellar photons. This process can dominate the $\gamma$-ray emission as well, competing with proton-proton inelastic collisions (p-p). Some $\gamma$-ray photons with energy $\gtrsim 100$~GeV can be absorbed in the stellar radiation field creating secondary electron-positron pairs. In our model we also calculate the wind thermal emission ---relevant at high radio frequencies--- using the standard expressions given by \cite{Wright1975} for a spherically symmetric stellar wind.

\section{RESULTS AND DISCUSSION}
\label{sec:results}

\begin{figure}[htpb]
\includegraphics[angle=270, width=\linewidth]{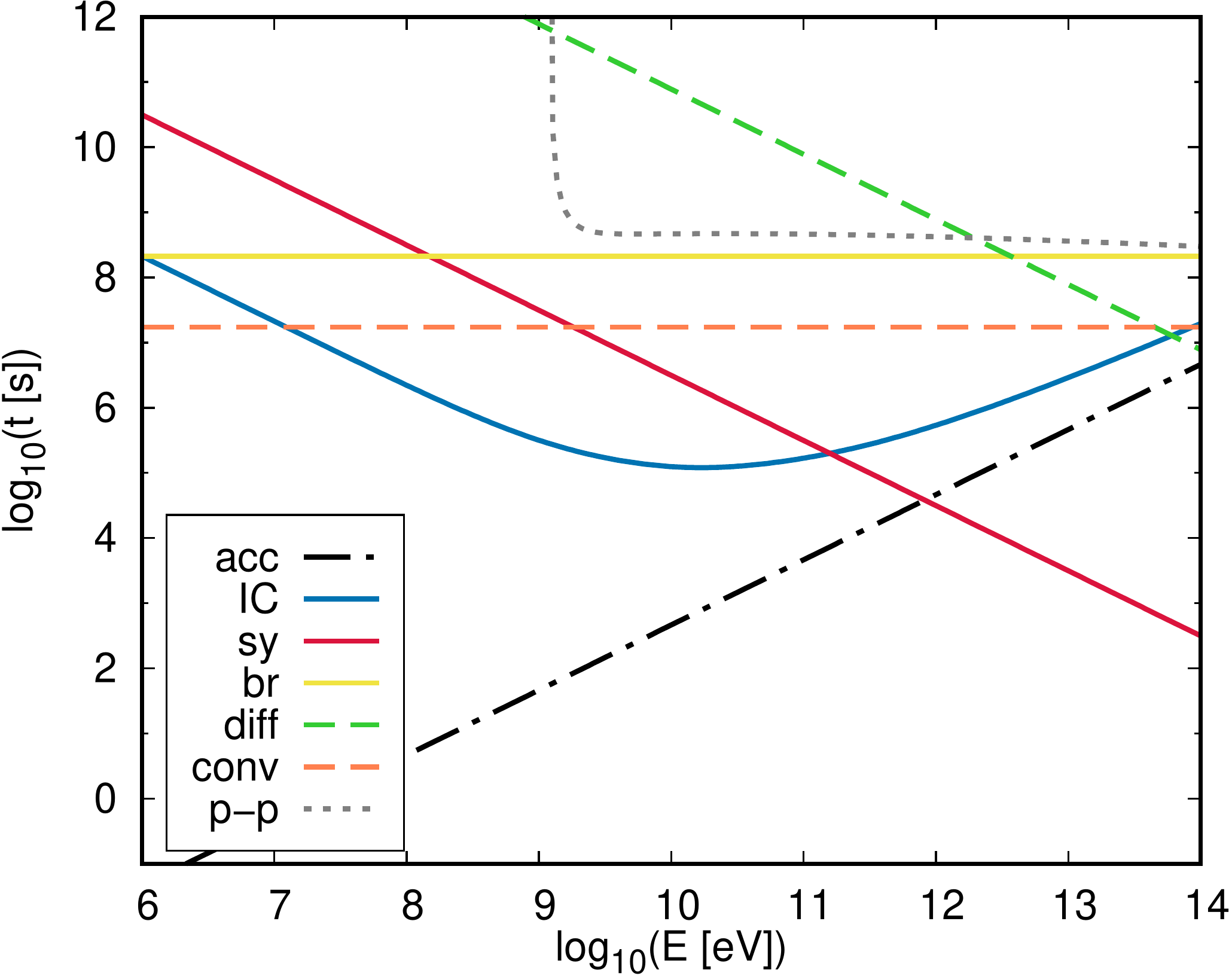}
\caption{Example of characteristic timescales $t$ for relativistic particles of energy $E$. These are calculated at the shock associated to the WN star in a position close to the apex of the WCR, assuming $\eta_B = 0.01$. The acceleration timescale is shown with a dash-dotted line, electron radiative cooling timescales for IC scattering with stellar photons (from both stars), synchrotron, and relativistic Bremsstrahlung are shown with solid lines, and the diffusive and convective escape timescales are shown in dashed lines. The cooling time for protons (p-p) is shown in a dotted line.}
\label{fig:t_char}
\end{figure}

The non-thermal emission model considered here takes into account the inhomogeneity of the emitter and the variable conditions along the shocks. We can, however, make a simplified discussion of the main properties that shape the particle energy distribution and the associated emission. In Fig.~\ref{fig:t_char} we show an example of the characteristic timescales for relativistic particles at a position close to the apex of the WCR for a given value of $\eta_B = 0.01$; for other locations in the emitter and other values of $\eta_B$ the overall behaviour is analogue. For the parameters of the Apep system, the electron energy distribution is governed by convective escape for $E_\mathrm{e} < 100$~MeV, whereas for $E_\mathrm{e} > 100$~MeV the IC losses become dominant. Nonetheless, the IC interactions have a transition from the Thomson to the Klein-Nishina regime at $E_\mathrm{e} \gtrsim 10$~GeV, thus making this mechanism less efficient as the cross-section of the interaction drops. Depending on the magnetic field intensity, synchrotron cooling can overcome the IC losses at $E_\mathrm{e} \gtrsim 100$~GeV. The cooling of the high-energy electrons leads to a softening of the energy distribution at GeV energies and also to a slight increase in the distribution of lower-energy electrons, thus changing the overall shape of the emitted SED. In the case of protons, cooling is negligible and the shape of their energy distribution is given by escape losses (Fig.~\ref{fig:t_char}); except for the highest energies ($\gtrsim10$~TeV), the escape is convective and the energy distribution has the same shape as the injected spectrum.
Finally, we note that 
the shock of the WN star wind is more powerful than the one from the WC star only by $\sim 20\%$, so that both stellar wind shocks contribute to the observed fluxes. 

\subsection{Radio emission maps}\label{sec:model_maps}

Given the axi-symmetry of the WCR at the spatial scales of interest, the morphology of the emission maps depends almost exclusively on the system orientation. We compute synthetic emission maps for different observing angles $\psi$. We convolve these maps with a Gaussian beam of $5.6'' \times 11.3''$, and normalise the simulations in such a way that the integrated flux density in the map matches the flux density reported by \cite{Marcote2021}, that is, $S_{2\,\mathrm{GHz}} = 60$~mJy~\footnote{We note that, as pointed out by \cite{Marcote2021}, this is not the total flux density at 2~GHz, which should be closer to 120~mJy, but rather the retrieved one from the interferometric observations. We take this into account by multiplying the map fluxes by a renormalisation factor of $60/120 = 0.5$.}. 

\begin{figure*}[htpb]
\centering
\includegraphics[angle=0, width=.9\textwidth]{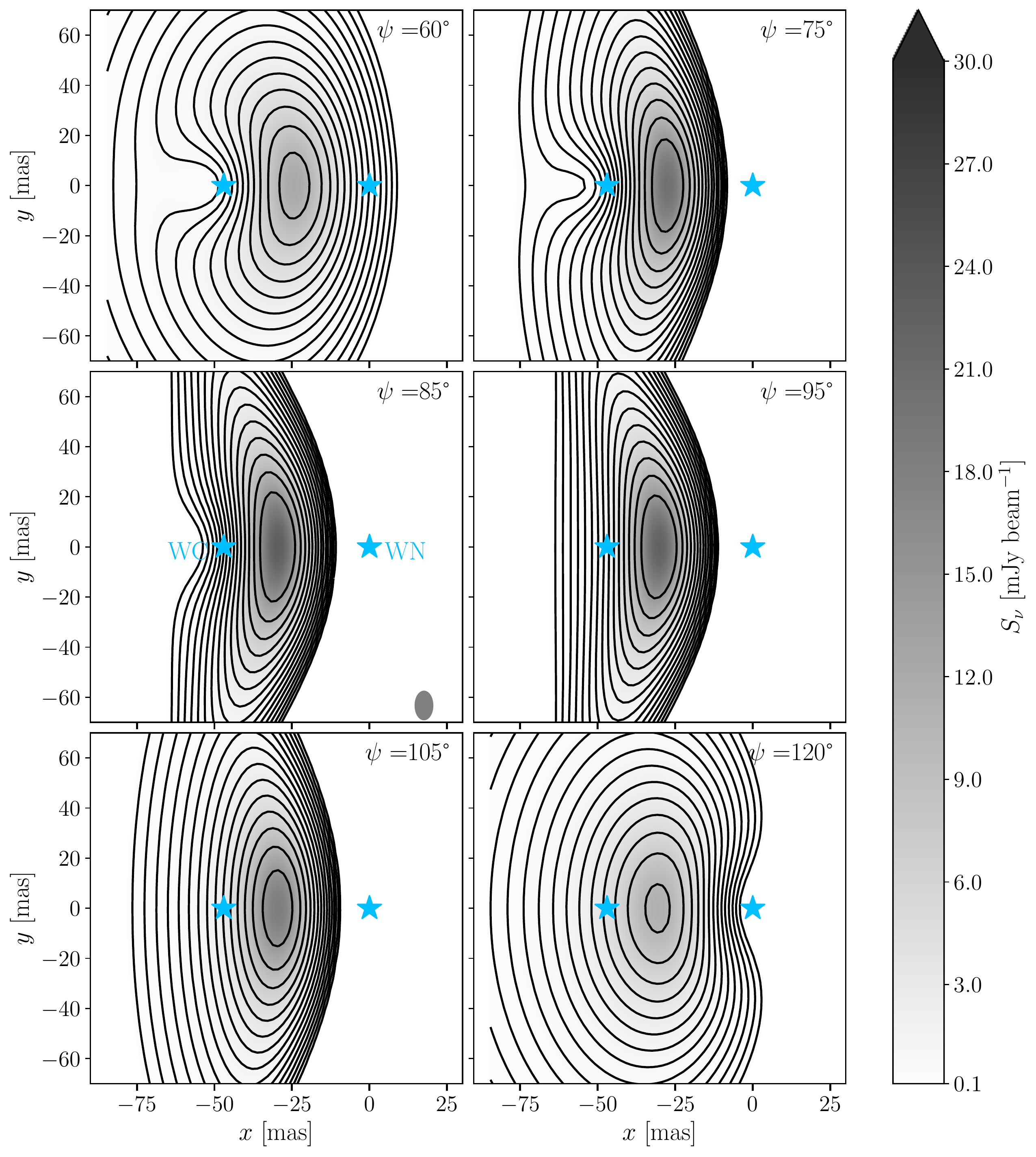}
\caption{Synthetic emission maps using the model described in Sec.~\ref{sec:model}. The position of the WC (left) and WN (right) stars are shown, together with the synthesised beam in the bottom right corner of the middle left panel. The intensity of the wind-collision region emission is shown in grayscale, and we overplot the same contour levels as in \cite{Marcote2021}. The maps match well with the observed morphology for $\psi \sim 85^\circ$.}
\label{fig:maps}
\end{figure*}

In Fig.~\ref{fig:maps} we show the synthetic emission maps. For $\psi < 90^\circ$, the WC star is in front, whereas for $\psi> 90^\circ$ the WN is in front. Despite the fact that the measurement by \cite{Marcote2021} is resolving out flux, we can constrain the value of $\psi$ comparing by eye the synthetic maps with the observed one. Values of $\psi < 65^\circ$ and $\psi > 105^\circ$ lead to emission much more spatially diluted than the observed one, as can be appreciated from the contour levels and/or the colorscale. For this reason we rule out these values, leaving us with a tentative value of $\psi = 85^\circ \pm 20^\circ$. Additionally, the case with $\psi = 85^\circ$ shows the best agreement with the observed map, and we therefore fix this value hereafter. We note, however, that for the range of values of $\psi = 65^\circ-105^\circ$, the distance between the stars varies in less than 10\%, and therefore the specific value adopted for $\psi$ has little impact in the results obtained in the following sections.

For consistency, we check whether this value is compatible with independent measurements by other authors. In particular, by modelling the IR spiral plume, \cite{Han2020} found the following orbital parameters for the Apep system: inclination $i=25^\circ \pm 5^\circ$, argument of periastron $\omega=0^\circ \pm 5^\circ$, and true anomaly at 2018 epoch $\nu=-173^\circ \pm 15^\circ$. Similar values (within errors) were also obtained by \cite{Bloot2021} by modelling the lightcurve at radio wavelengths.
We can derive the corresponding projection angle for these parameters as $\psi~=~\arctan{( \sqrt{x^2 + y^2}/z )}$, where $x=D\cos{(\omega-\nu)}\cos{(i)}$, $y=D\sin{(\omega-\nu)}$, and $z=D\cos{(\omega-\nu)}\sin{(i)}$ are the coordinates of the secondary star. 
We obtain $\psi \approx 70^\circ$, which is roughly consistent with the value we derive independently from the morphology of the emission maps.

\subsection{Radio spectral energy distribution} \label{sec:model_radio}

\begin{figure}[ht]
\includegraphics[width=\linewidth]{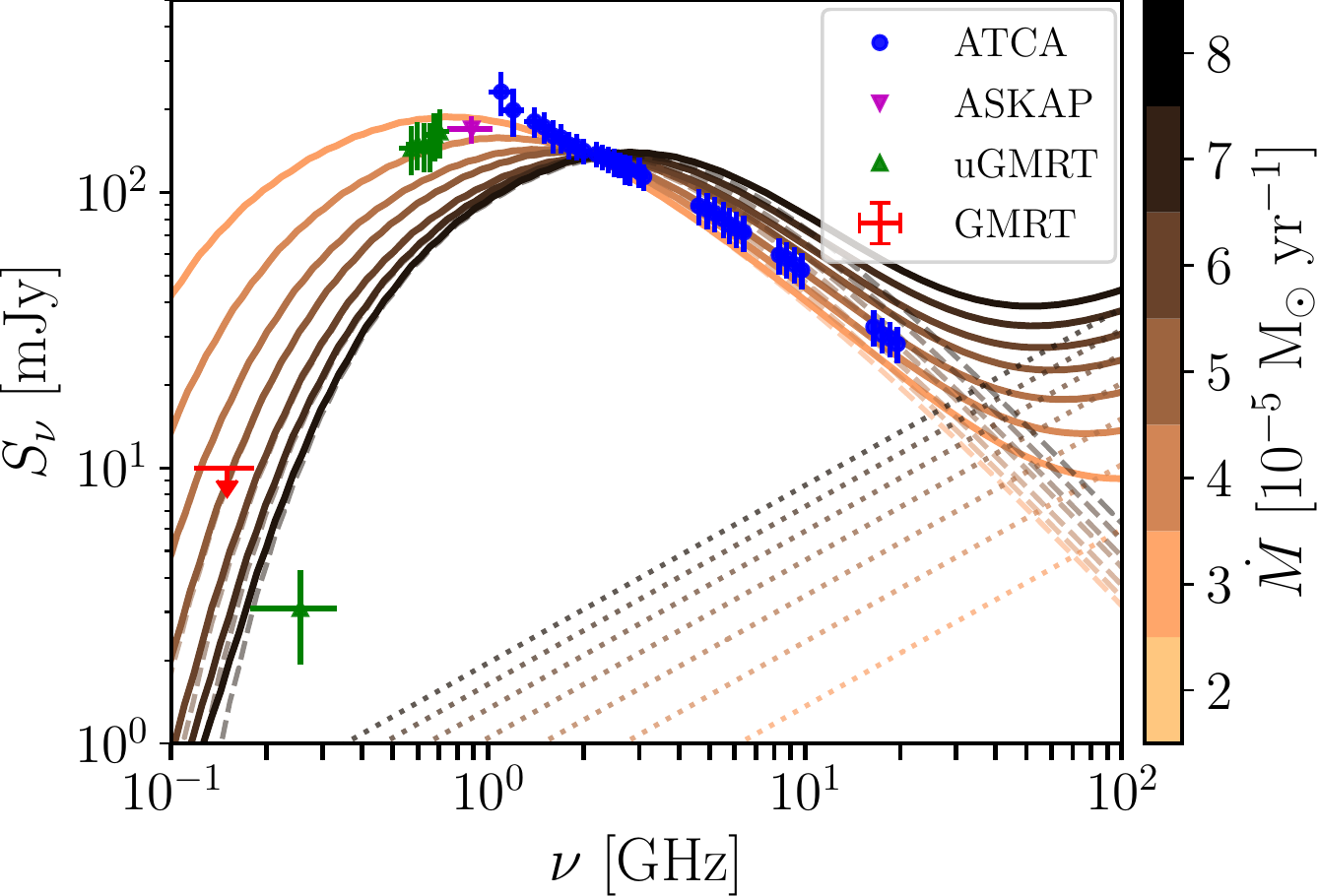}
\caption{Modelled SED of Apep at radio frequencies for different values of the WN star mass-loss rate, as indicated in the colorbar. We show the observational data points taken with the uGMRT and ATCA \citep{Bloot2021}, and the upper limit at 150~MHz calculated in this work from the GMRT 150~MHz all-sky radio survey \citep{Intema2017}. Dotted lines show the total free--free from the stellar winds, dashed lines the synchrotron component from the wind-collision region, and solid lines are the sum of both.}
\label{fig:seds_radio}
\end{figure}

The shape of the SED at radio frequencies depends strongly on the stellar mass-loss rates. This is because the spectrum below 1~GHz is severely affected by FFA in the ionised stellar winds. Moreover, the thermal free--free emission is also relevant at frequencies above 10~GHz. Both features are more pronounced for higher values of $\dot{M}$. We therefore carry out simulations for different values of $\dot{M}_\mathrm{WN}$ (and, consistently, of $\dot{M}_\mathrm{WC}$; see Sec.~\ref{sec:apep}) and compute the radio SED in each case. All the SEDs are normalised such that $S_{2\,\mathrm{GHz}} = 120$~mJy; for doing this, we fix $f_\mathrm{NT} = 0.1$ ($f_\mathrm{NT,e} = 0.005$) and vary the magnetic field intensity in the WCR through the parameter $\eta_B$. For the cases considered of $\dot{M}_\mathrm{WN} = (2-8)\times10^{-5}$~M$_\odot$\,yr$^{-1}$, the $\eta_B$-values are in the range $(0.17-1.1)\times 10^{-2}$, with the lower values of $\eta_B$ corresponding to the greater values of $\dot{M}_\mathrm{WN}$. 
We show our results in Fig.~\ref{fig:seds_radio}, together with the available observational data points.

The available radio data allow us to put strong constraints on the value of $\dot{M}_\mathrm{WN}$. The flux densities at 1--3~GHz allow us to characterise the intensity of the synchrotron spectrum, while the flux densities and upper limit below 1~GHz help us to infer the position of the FFA turnover frequency. Moreover, the flux density value at 19.7~GHz further constrains the combination of the synchrotron SED and the thermal free--free emission from the winds. 

According to the results shown in Fig.~\ref{fig:seds_radio}, the shape of the radio SED cannot be reconciled with low mass-loss rates, $\dot{M}_\mathrm{WN} < 3\times10^{-5}$~M$_\odot$\,yr$^{-1}$, as the flux density below 300~MHz would be highly overpredicted. In addition, high values of $\dot{M}_\mathrm{WN} > 5\times10^{-5}$~M$_\odot$\,yr$^{-1}$ lead to a significant absorption up to $\sim 2$~GHz, in tension with the data at 0.6--1.4~GHz, and they also lead to an overestimation of the total flux density at 19.7~GHz. Therefore, we conclude that the value of $\dot{M}_\mathrm{WN}$ is well-constrained to the range $(3-5)\times10^{-5}$~M$_\odot$\,yr$^{-1}$, which leads to $\dot{M}_\mathrm{WC} \approx (2.2-3.7)\times10^{-5}$~M$_\odot$\,yr$^{-1}$. For these mass-loss rates the model reproduces quite well the overall shape of the radio SED, but not the very pronounced decline suggested by the data at 255~MHz. Hereafter we adopt a reference value of $\dot{M}_\mathrm{WN} = 4 \times10^{-5}$~M$_\odot$\,yr$^{-1}$. 

Nonetheless, we caution that these estimates are subject to larger uncertainties when taking into account the uncertainties in the wind parameters, particularly in those adopted as typical values for high-mass stellar winds (Table~\ref{tab:parameters}). For instance clumping yields $\dot{M} \propto f^{-1/2}$, so that variations in the assumed value of $f$ within a factor of two would yield variations in the estimated values of $\dot{M}$ within a factor $\sqrt{2}$. Additional uncertainties come from other system parameters, such as the distance $d$ to the source, although in this case the values of $\dot{M}$ vary approximately by only  $\Delta d/d < 20\%$. 

In addition, \cite{Callingham2019} found observational evidence of anisotropy in the stellar winds, which is a feature not included in our model. Moreover, \cite{Bloot2021} also favoured the presence of anisotropic winds in Apep by means of modelling the radio lightcurve using a one-zone approximation for the emitter. However, such a one-zone model treats the emitter as a point-like homogeneous region and is therefore incapable of accounting for the extended nature of the WCR and the fact that the emitted photons probe different regions of the stellar winds depending on their production site \citep[e.g.][]{Dougherty2003}. Thus, we expect that the actual impact of the wind anisotropies in the radio SED to be less significant than implied by \cite{Bloot2021}. With respect to the spherical wind approximation in our model, depending on the geometry of the system and the winds anisotropy, the anisotropy in the winds could potentially increase the opacity in the direction of the line of sight for photons coming from close to the bright apex of the WCR. Such a possibility could help to relieve the tension with the data at 255~MHz. 

Finally, we note that the observations by \cite{Marcote2021} only revealed the WCR, but they did not detect the individual stars. Therefore, we cannot rule out the possibility that the positions of the WR stars are exchanged. This means that in Fig.~\ref{fig:maps} the WC could actually be the star to the right, having the strongest wind. In this scenario, the value of $\eta = 0.44$, together with the wind terminal velocities, yield $\dot{M}_\mathrm{WC} \approx 4\dot{M}_\mathrm{WN}$. This leads to some complications, however. First, as shown previously, higher mass-loss rates can hardly be reconciled with the observed SED at high frequencies and with the non-detection of the individual WR stars by \cite{Marcote2021}. One would actually need to reduce $\dot{M}_\mathrm{WN}$ to achieve the above constraint. This, in turn, would lead to lower absorption in the radio SED, which again is inconsistent with the observations at low frequencies. In addition, the lower wind kinetic power would demand an even more efficient conversion of wind kinetic power into non-thermal particles. We conclude that the adopted configuration, with the WN star to the right (i.e., having the strongest wind), seems to be the most plausible choice.

\subsection{High-energy spectral energy distribution}
 
Having solved the uncertainties in $\psi$ and $\dot{M}$ to a great extent, we now focus on the degeneracy of the two remaining free parameters in the model: $\eta_B$ and $f_\mathrm{NT,e}$. Such a degeneracy cannot be solved by radio data alone, but predictions in the high-energy domain can help to break this degeneracy \citep[e.g.][]{delPalacio2020}. Here we calculate the expected fluxes in different energy bands accessible to current facilities. In particular, we report the fluxes in the hard X-ray energy range observable with the satellite \textit{NuSTAR} ($10$--$79$~keV), the $\gamma$-ray energy ranges observable by \textit{Fermi}-LAT ($0.1$--$100$~GeV) and the forthcoming Cherenkov Telescope Array (CTA; $0.1$--$100$~TeV). We also compare the predicted fluxes in the MeV band with the expected sensitivity of the \textit{e-ASTROGAM} mission \citep{eASTROGAM2018}.

\begin{figure}[htpb]
\includegraphics[width=\linewidth]{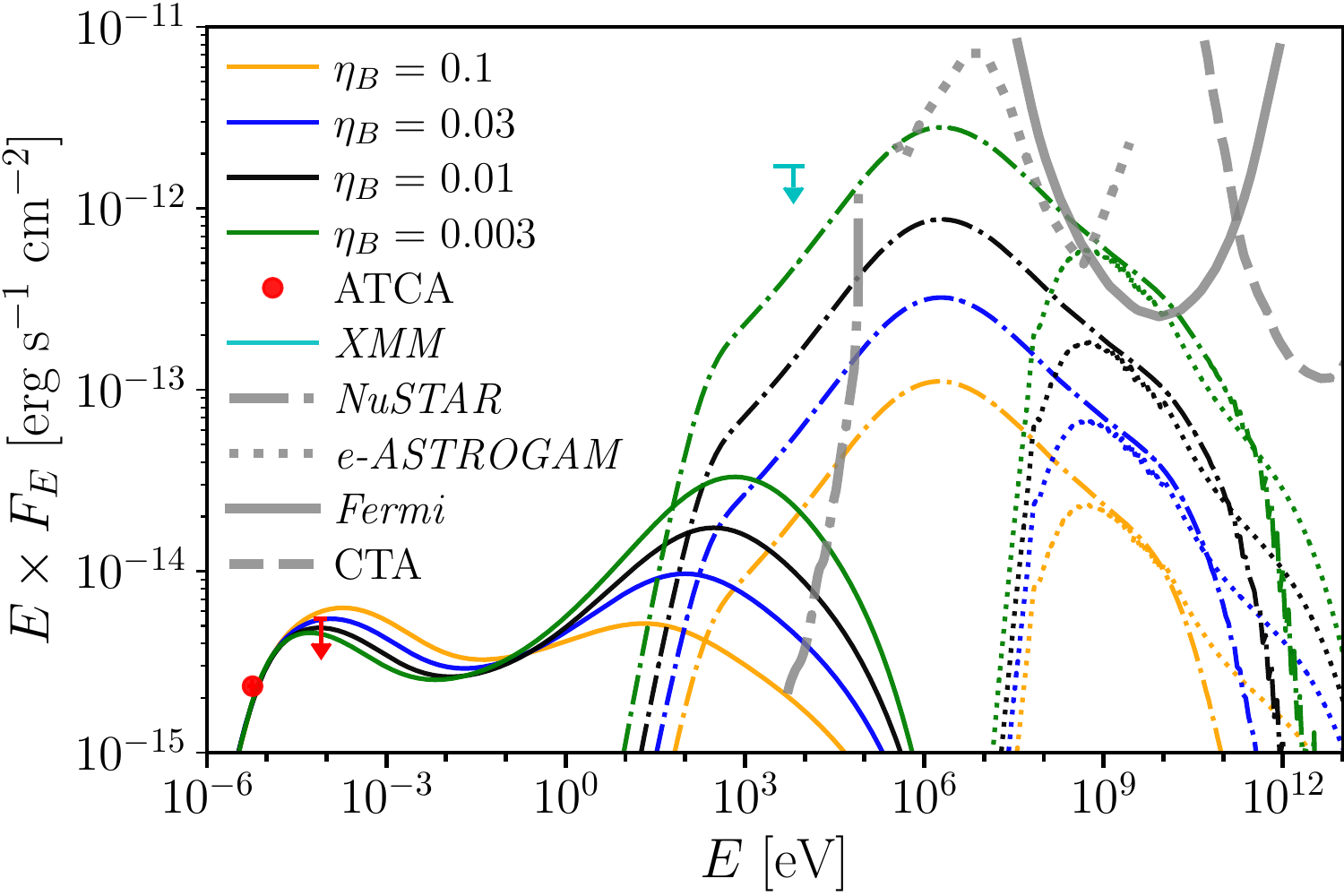}
\caption{Modelled non-thermal SED of Apep for different values of $\eta_B$. We show the observational data points taken with ATCA \citep{Callingham2019} and the upper limit we derive from \textit{XMM-Newton} data (Sec.~\ref{sec:x-ray_obs}). Dotted lines show the p-p component, dot-dashed lines the IC component, and solid lines the synchrotron component from the wind-collision region; all emission components are absorption-corrected. We also show the sensitivity curves for 1-Ms \textit{NuSTAR} \citep{Koglin2005}, 1-yr \textit{e-ASTROGAM} \citep{eASTROGAM2018}, 10-year \textit{Fermi}-LAT (extracted from \url{https://www.slac.stanford.edu/exp/glast/groups/canda/lat_Performance.htm} for a broadband detection)
and 100-h CTA \citep[extracted from][]{Funk2013}.}
\label{fig:seds_etas}
\end{figure}

We calculate the broadband SED for different scenarios that cover plausible physical conditions at the shocks. We explore different values of $\eta_B$ and fit the fraction of the available wind kinetic power transferred to relativistic electrons in the shocks needed to match the observed flux density at 2~GHz. For each of these scenarios we also calculate the IC emission produced by the same population of relativistic electrons, and the p-p emission produced by the relativistic protons. The results are shown in Fig.~\ref{fig:seds_etas}. We note that the synchrotron and IC SED deviate from a simple power law due to the efficient IC cooling of the high-energy electrons.

If we consider a case of high magnetic field given by $\eta_B~=~0.1$ ($B_\mathrm{WCR} \sim0.4$~G), which is relatively close to the pressure equipartition condition, we obtain a corresponding value of $f_\mathrm{NT} = 5.4\times 10^{-3}$. However, the strong magnetic field enhances the synchrotron emission from low-energy electrons that emit at $\nu \gtrsim 10$~GHz, surpassing the detected flux density at 19.7 GHz by a $\sim 30\%$; this leads to a significant tension considering that the stellar winds should also have a relevant contribution at these frequencies (see Sec.~\ref{sec:radio_obs}). If we instead consider a low magnetic field case with $\eta_B = 0.001$ ($B_\mathrm{WCR} \sim0.04$~G), we obtain a high value of $f_\mathrm{NT} = 0.42$ ($f_\mathrm{NT,e} = 0.081$), which yields an IC flux in the 3--10~keV band of $\gtrsim 9\times10^{-13}$~\flux that is in tension with the upper limits set by X-ray observations (Sec.~\ref{sec:x-ray_obs}). We can therefore rule out models with $\eta_B < 0.002$ as they overpredict the X-ray flux in the 3--10~keV band (Sec.~\ref{sec:x-ray_obs}).
In conclusion, we can constrain $0.002~<~\eta_B~\leq~0.1$. Below, we present a more detailed analysis of the cases $\eta=0.003$, $\eta=0.01$, $\eta=0.03$ and $\eta=0.1$.

For the values of $\eta_B \approx 0.003-0.1$ ($B_\mathrm{WCR} \sim 0.08-0.4$~G) we derive corresponding values of $f_\mathrm{NT} \approx 0.005-0.13$ ($f_\mathrm{NT,e} \approx (0.11-2.7)\times 10^{-3}$), which can vary by a factor $0.85-1.6$ when taking into account the uncertainty in the distance to the system. These values are consistent with those found for the CWB HD~93129A by \cite{delPalacio2020} ($\eta_B\approx0.02$, $B_\mathrm{WCR} \sim 0.5$~G, and $f_\mathrm{NT,e}\approx6\times10^{-3}$). 
Moreover, we show that the $\gamma$-ray emission comes from a combination of IC radiation and p-p interactions. However, the predicted fluxes are below the sensitivity of current $\gamma$-ray facilities by an order of magnitude, except in the most favourable scenarios (Fig.~\ref{fig:seds_etas}). This is mainly due to the rather soft spectral index of the particle energy distribution ($p=2.42$). In addition, for photon energies $\sim 100$~GeV, $\gamma$--$\gamma$ absorption is relevant and diminishes the flux of the source by $\approx 25\%$. Nonetheless, a possible hardening of the particle energy distribution at high energies could potentially increase the $\gamma$-ray luminosity significantly \citep[e.g.][]{delPalacio2016, delPalacio2020}. We also note that the radiation from secondary pairs created in $\gamma$--$\gamma$ interactions is negligible because the soft $\gamma$-ray spectrum leads to much less power being radiated above 100 GeV than below 100 GeV, so the pair energetics is small. We summarise the predicted fluxes for each scenario in Table~\ref{tab:fluxes}.

\begin{table*}
\caption{Fluxes in different energy bands for different values of $\eta_B$. The selected energy bands correspond to the ones accessible by \textit{NuSTAR} (3--79~keV), \textit{Fermi} (0.1--100~GeV), and imaging air Cherenkov telescopes such as CTA (0.1--100~TeV).}
\centering
\begin{tabular*}{\textwidth}{@{}c\x c\x c\x c\x c\x c@{}}
\hline \hline
 $\eta_B$   & $f_{\rm NT}$   & $F_{\rm 3-10 keV}$  &  $F_{\rm 10-79 keV}$  &  $F_{\rm 0.1-100 GeV}$ & $F_{\rm 0.1-100 TeV}$ \\
 ~  & [$10^{-2}$]       &     [$10^{-14}$\,\flux]         &     [$10^{-13}$\,\flux]        &   [$10^{-10}$\,\flux]  &   [$10^{-14}$\,\flux]  \\
\hline
 0.1    & 0.54   & 1.2  & 0.4  & 1.6  & 0.8 \\ 
 0.03   & 1.56   & 3.3  & 1.2  & 4.9  & 3.0 \\ 
 0.01   & 4.2    & 8.9  & 3.1  & 13.1 & 9.5 \\ 
 0.003  & 13.5   & 28.5 & 10.1 & 42.4 & 36.2 \\ 
\hline \hline
\end{tabular*}
\label{tab:fluxes}
\end{table*}

\section{CONCLUSIONS}
\label{sec:conclusions}

We present a detailed study of the non-thermal emission from the CWB Apep. The main results from this work are:

\begin{itemize}

    \item We constrain the observing projection angle to be $65^\circ \leq \psi \leq 105^\circ$, with $\psi \sim 85^\circ$ the preferred value, by modelling the morphology of the emission maps. 
    
    \item We establish upper limits of the non-thermal radio emission at 150~MHz of 10~mJy and of the non-thermal X-ray emission in the 3--10~keV energy band of $1.7\times10^{-12}$~\flux.
    
    \item We estimate the stellar mass-loss rate of the WR stars to be $\dot{M}_\mathrm{WN} \approx (4 \pm 1)$~\Msunyr and $\dot{M}_\mathrm{WC} \approx (2.9 \pm 0.7)$~\Msunyr by modelling the radio SED (for a wind volume filling factor $f \approx 0.2$).
    
    \item We constrain the magnetic field intensity in the WCR and the fraction of energy converted into non-thermal particle acceleration. Namely, these values are $\eta_B \approx 0.003-0.1$ ($B_\mathrm{WCR} \approx 0.08-0.4$~G) and $f_\mathrm{NT,e} \approx (0.11-2.7)\times 10^{-3}$.

    \item We predict the expected emission of the Apep system at high energies (hard X-rays and $\gamma$-rays). The CWB Apep is unlikely to be detected as a $\gamma$-ray source unless the particle energy distribution has a hardening at high energies. 

\end{itemize}

We conclude that the detailed investigation of the non-thermal radiation from CWBs such as Apep can offer deep insights on the general picture of CWBs, both as particle accelerators and non-thermal sources.


\begin{acknowledgements}
S.d.P. acknowledges support by CONICET (PIP-0102) and ANPCyT (PICT-2017-2865); P.B., by ANPCyT (PICT-2017-0773).
G.E.R. \& V.B-R. are supported by the Spanish Ministerio de Ciencia e Innovaci\'{o}n (MICINN) under grant PID2019-105510GB-C31 and through the ``Center of Excellence Mar\'{i}a de Maeztu 2020-2023'' award to the ICCUB (CEX2019-000918-M). V.B-R. is also supported by the Catalan DEC grant 2017 SGR 643, and is Correspondent Researcher of CONICET, Argentina, at the IAR. We thank H. Intema for help with TGSS ADR1 data handling. This work was carried out in the framework of the PANTERA-Stars\footnote{\url{https://www.astro.ulg.ac.be/~debecker/pantera/}} initiative.
\end{acknowledgements}

\begin{appendix}

\section{PARTICLE ENERGY DISTRIBUTION}\label{appendix:dist}

The WCR is treated as a sum of one-dimensional emitters. Each of these is divided into smaller segments or ``cells''. The relativistic particle distribution from one cell evolves as it reaches the next cell. To calculate this evolution, one has to take into account both the energy losses and the particle travel time along each cell, which depends on the cell size and the fluid velocity \citep[e.g.][]{Molina2018}. Specifically, the particle energy distribution evolves as 
\begin{equation}
        N(E', i+1) = N(E, i) \frac{ |\dot{E}(E,i)|}{ |\dot{E}(E,i+1)|} \frac{t_\mathrm{cell}(i+1)}{t_\mathrm{cell}(i)},
       \label{Eq:N(E)}
\end{equation}
where $\left|\dot{E}(E, i)\right|$ is the cooling rate for particles of energy $E$ at the $i$-cell. In the base model by \cite{delPalacio2016}, the factor $t_{\rm cell}$ was not included, which can lead to errors in the estimated fluxes of $\sim 10\%$ depending on the cell sampling adopted and the fluid acceleration. 

\end{appendix}


\bibliographystyle{pasa-mnras}
\bibliography{biblio}

\end{document}